\documentstyle[12pt,xy]{article}
\xyoption{v2}    
\title{On the law of motion in Special Relativity}
\author{Gonzalo E. Reyes \\ D\'{e}partement de math\'{e}matiques \\
Antoine Royer \\ \'{E}cole Polytechnique \\
Universit\'{e} de Montr\'{e}al}
\date{February 2003}

\font\zbb=bbold12 

\def\bb#1{\mbox{\zbb #1}}

\def\bf#1{\mbox{\boldmath$#1$}}
\begin{document} 
\maketitle 

\section*{Abstract}
Newton's law of motion for a particle of mass $m$ subject to a force
$\bf f$ acting at time $t$ may be formulated either as
$$
\bf f=d/dt\; (m\bf u(t))
$$
or, since $m=m_0$ is a constant,  as
$$
\bf f=m\bf a(t)
$$
where $\bf u(t)$ and $\bf a(t)$ are the velocity and the acceleration, respectively, of the particle at time $t$ relative to an inertial frame $S,\;$ `the laboratory'. This law may be interpreted in either of two ways:
\begin{enumerate}
\item[(1)] The force \bf f acting on the particle at time $t$ during an infinitesimal time $\delta t$ imparts to the laboratory a boost $ \delta \bf u=(1/m)\bf f \delta t,$ while the particle maintains the velocity $\bf u(t)$ relative to the new frame $S'$. 
\item[(2)] The force \bf f acting on the particle at time $t$ during an infinitesimal time $\delta t$ imparts to the particle a boost $\delta\bf w=(1/m)\bf f \delta t$ relative to its proper frame $S_0$ which moves with velocity $\bf u(t)$ relative to $S$.
\end{enumerate}
We show that the relativistic law of motion admits both interpretations, the first of which is in fact equivalent to the law of motion. As a consequence, we show that the relativistic law of motion may also be reformulated as 
$$
\bf f=m\bf a
$$ 
in analogy with Newton's law, but with a {\it relativistic} mass and a {\it relativistic} acceleration defined in terms of the {\it relativistic} addition law of velocities, rather than ordinary mass and ordinary vectorial addition of velocities that lead to the classical acceleration and to Newton's law. 

\section*{Introduction}
It is well-known that the Special Theory of Relativity is based on two postulates: the principle of relativity and the constancy of the speed of light. From these postulates, one can deduce the Lorentz transformations that connect the space-time coordinates of a particular event  in two inertial frames.  From these transformations, in turn, lenght contraction, time dilation, the `addition' formula for velocities, etc., follow straightforwardly (see e.g \cite{ein}, \cite{mo}, \cite{pau}, \cite{rin}). Thus, these postulates are sufficient for the  development of the theory as far as {\it kinematics} is concerned. 

\medskip
On the other hand, this is not so for {\it dynamics}. In the earlier papers, the law of motion was obtained from electrodynamical considerations. After a tortuous path, Planck and independently Tolman, finally arrived at the familiar formulation of today connecting force, mass and velocity at a given time:
$$\left \{ \begin{array}{c}
\bf f=d/dt(m\bf u(t))\\ 
m=m_0\gamma (u)\\
\gamma (u)=1/\sqrt{(1-u^2/c^2)}
\end{array}
\right.$$  
The time $t$, the mass $m$, the force \bf f and the velocity $u(t)$ are relative to an inertial frame $S$. (See  \cite{dug}, \cite{ein},  \cite{aim} and \cite{pau} for historical references).
 
\medskip
In his lecture on `Space and Time' (see \cite{ein}), delivered in 1908 and published a year later, Minkowski introduced an {\it invariant} reformulation of the usual law of motion: 
$$\bf F=m_0\bf A$$
where $\bf F$ is the 4-force acting on the particle and $\bf A$ the 4-acceleration.

\medskip
It should be noticed, however, that $\bf F$ is a 4-force and thus differs from the 3-force $\bf f$. Even the spatial component $(F_1, F_2, F_3)$ of Minkowski 4-force differs from the previously introduced 3-forces. In fact
$$ (F_1, F_2, F_3)={\gamma}(u) \bf f$$
The corresponding spatial component of the 4-acceleration is therefore
$$ (A_1,  A_2, A_3) ={\gamma}(u)d/dt({\gamma}(u)\bf u)$$
The time components of the Minkowski force and the acceleration are given by
$ F_4=(\gamma (u)/c) \bf f.\bf u$ and $ A_4=c\gamma (u)d/dt(\gamma (u)).$ The equation $ F_4=m_0 A_4$ follows from the corresponding equation for the spatial component and is in fact equivalent to $\bf f.d\bf r=c^2dm,$ as shown by an easy calculation.

\medskip
The arguments generally given in favor of the usual law of motion are (see \cite{rin2}):
\begin{enumerate}
\item[(i)] it reduces to the law of Newton $\bf f= m\bf a$ in the classical limit
\item[(ii)] it leads to the conservation of momentum in simple collisions, provided that we assume the law of equality of action and reaction at contact
\item[(iii)] it leads to $\delta E=c^2\delta m,$ the infinitesimal version of Einstein's famous law $E=mc^2$ by adopting the classical definition of work as $force\times distance,$ as we will see later on.
\item[(iv)] it is consistent with the well-established Lorentz law of force in electrodynamics
\end{enumerate}
These are strong arguments not only in favor of the usual law of motion, but (iii) and (iv) provide good reasons to consider also the 3-force (along with the Minkowski 4-force ).

\medskip
The aim of this note is to show that this law of motion admits the two interpretations of Newton's law stated in the Abstract. To formulate them in a concise manner, we shall use the diagram 
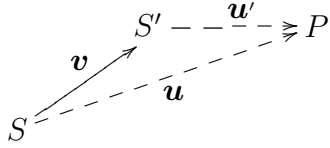
\begin{center}
\leavevmode
\xy\xymatrix{
&\;\;\; S'--\arrow @{-->} [r]^{\bf u'}  &P \\
S \arrow @{->} [ru]^{\bf v} \arrow @{-->} [rru]_{\bf u} &\\
}
\endxy
\end{center}
as a shorthand to describe the following situation: the particle $P$ moves with velocity $\bf u'$ relative to $S',$  which itself moves with velocity $\bf v$ relative to $S$ and $\bf u$ is the velocity of the particle relative to $S.$ It will be particularly handy when we deal with relativity and Lorentz transformations.

\medskip
To simplify the notation, we shall use either $\partial_t$ or $\dot{(...)}$ for $d/dt$. i.e., the derivative with respect to time.

\section*{Newtonian Dynamics}
Newton's law of motion for a particle of mass $m$ subject to a force
$\bf f$ acting at time $t$ is
$$
\bf f=\partial_t (m\bf u(t))
$$
where $\bf u(t)$ is the velocity of the particle at time $t$ relative to an inertial system $S,$ to be referred to as `the laboratory'.

\medskip
This law can be rewritten as 
$$\bf u(t+\delta t)=\bf u(t)+(1/m)\bf f\delta t $$
by letting $\delta t$ an infinitesimal time increment and using 
$$
\bf u(t+\delta t)=\bf u(t)+ \dot{\bf u}(t)\delta t
$$
 Now, the point is that this law may be interpreted in either of two equivalent ways:
\begin{enumerate}
\item[(1)] The force \bf f acting on the particle at time $t$ during an infinitesimal lapse of time $\delta t$ imparts to the laboratory $S$ , relative to which the particle has velocity \bf u(t), a boost $ \delta \bf u=(1/m)\bf f \delta t$ while the particle maintains the velocity $\bf u(t)$ relative to the new frame $S'$. Thus, the new velocity of the particle relative to $S$ at time $t+\delta t$ is
$$
 \bf u(t+\delta t)=(1/m)\bf f \delta t + \bf u(t)
$$
Using our diagram,
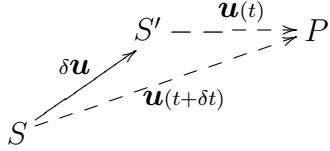
\begin{center}
\leavevmode
\xy\xymatrix{
&\;\;\; S'--\arrow @{-->} [r]^{\bf u(t)}  &P \\
S \arrow @{->} [ru]^{\delta \bf u} \arrow @{-->} [rru]_{\bf u(t+\delta t)} &\\
}
\endxy
\end{center}
\item[(2)] The force \bf f acting on the particle at time $t$ during an infinitesimal lapse of time $\delta t$ imparts to the particle a boost $\delta\bf w=(1/m)\bf f \delta t$ relative to its proper frame $S_0$ which moves with velocity $\bf u(t)$ relative to $S$. Then the new velocity of the particle relative to $S$ at time $t+\delta t$ is 
$$
\bf u(t+\delta t)=\bf u(t)+(1/m)\bf f\delta t
$$ 
Diagrammatically, 
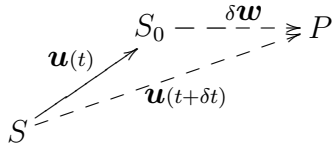
\begin{center}
\leavevmode
\xy\xymatrix{
&\;\;\; S_0--\arrow @{-->} [r]^{\delta \bf w}  &P \\
S \arrow @{->} [ru]^{\bf u(t)} \arrow @{-->} [rru]_{\bf u(t+\delta t)} &\\
}
\endxy
\end{center}
\end{enumerate}

\section*{Relativistic Dynamics}
As we mentioned already, in Special Relativity  Newton's law of motion is replaced by
$$\left \{ \begin{array}{c}
\bf f=\partial_t(m\bf u(t))\\ 
m=m_0\gamma (u)
\end{array}
\right.$$  
From now on, we choose a system of unities such that $c=1$. Thus
$\gamma (u)=1/\sqrt{1-u^2}$.

\medskip 
We wish here to point out a suggestive interpretation of the law of motion which does not seem to have been noticed before. In fact, we wish to point out that this law admits precisely the two previous interpretations, {\it provided} that we take care to refer all quantities to the corresponding frames, since force, mass and time are frame-dependent, contrary to the classical case, and take into account the fact that relativistic composition of velocities $\oplus$ is {\it not commutative}. Unlike the Newtonian case, however, only the first interpretation is equivalent to the law of motion. The second, although a consequence of this law, does not seem to be equivalent to it.

\medskip
Thus, 
\begin{enumerate}
\item[(1)] The force \bf f acting on the particle at time $t$ during an infinitesimal lapse of time $\delta t$ imparts to the laboratory $S$ , relative to which the particle has velocity \bf u(t), a boost $ \delta \bf u=(1/m)\bf f \delta t$ (where \bf f, $\delta t$ and m {\it are measured in $S$}), while the particle maintains the velocity $\bf u(t)$ relative to the new frame $S'$. Thus, the new velocity of the particle relative to $S$ at time $t+\delta t$ is
$$
 \bf u(t+\delta t)=(1/m)\bf f\delta t\oplus \bf u(t)
$$
Using our diagram once again, we obtain 
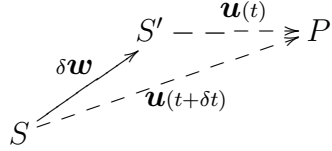
\begin{center}
\leavevmode
\xy\xymatrix{
&\;\;\; S'--\arrow @{-->} [r]^{\bf u(t)}  &P \\
S \arrow @{->} [ru]^{\delta \bf w} \arrow @{-->} [rru]_{\bf u(t+\delta t)} &\\
}
\endxy
\end{center}
\item[(2)] The force \bf f acting on the particle at time $t$ during an infinitesimal lapse of time $\delta t$ imparts to the particle a boost $\delta w_0=(1/m_0)\bf f_0(\delta t)_0$ relative to $S_0,$ its rest frame which moves with velocity $\bf u(t)$ relative to $S$. (Here \bf f, $(\delta t)_0$ and $m_0$ are measured in $S_0$). Then the new velocity of the particle relative to $S$ at time $t+\delta t$ (measured in $S$) is
$$
\bf u(t+\delta t)=\bf u(t)\oplus (1/m_0)\bf f_0(\delta t)_0
$$
Diagrammatically, 
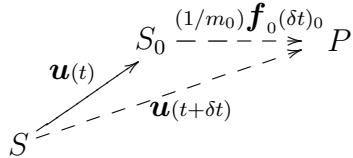
\begin{center}
\leavevmode
\xy\xymatrix{
&\;\;\; S_0 --\arrow @{-->} [r]^{(1/m_0)\bf f_0(\delta t)_0}  &\;\;P \\
S \arrow @{->} [ru]^{\bf u(t)} \arrow @{-->} [rru]_{\bf u(t+\delta t)} &\\
}
\endxy
\end{center}
\end{enumerate}
To prove our claim, we use some basic facts about relativistic kinematics that can be found in the Appendix.

\medskip
When the velocity  $ \bf v$ is infinitesimal, $\bf v=\delta \bf w,$ we get the infinitesimal Lorentz transformation
$$
\begin{array}{lll}
\delta \bf v\oplus \bf u &=& (\bf u+\delta \bf v)/(1+\delta \bf v.\bf u) \\
                                   &=& (\bf u+\delta \bf v)(1-\delta \bf v.\bf u) \\
                                   &=&  \bf u+\delta \bf v-\bf u(\bf u.\delta \bf v))
\end{array}
$$
since $\gamma(\delta \bf v)=1$

\medskip
Then 
$$
\bf u.\partial_t\gamma u = \bf u.(\dot{\gamma}\bf u+\gamma \dot{\bf u}) 
                                   = \dot{\gamma}u^2+\gamma \bf u.\dot{\bf u} 
                                   =  \dot{\gamma}u^2+\dot{\gamma}/\gamma^2 
                                   =  \dot{\gamma} \;\;\;\;\;(*)
$$
since $\dot{\gamma}=\gamma^3\bf u.\dot{\bf u}.$

\medskip 
Assuming the law of motion $\bf f=m_0\partial_t(\gamma \bf u)$,
$$
\delta w = \delta t\bf f/m 
                                   =\delta t\gamma^{-1}\partial_t\gamma \bf u 
                                   =  \dot{\bf u}\delta t+\delta t(\dot{\gamma}/\gamma)\bf u 
$$
and
$$
\bf u.\delta w = \bf u.\delta t\gamma^{-1}\partial_t\gamma \bf u 
                                   =\delta t\gamma^{-1}\bf u.\partial_t\gamma \bf u 
                                   =  \delta t(\dot{\gamma}/\gamma) 
$$
Thus, 
$$
\begin{array}{lll}
\delta \bf w\oplus \bf u&=& \bf u+\delta \bf w -\bf u(\bf u.\delta w) \\
                                   &=&  \bf u+\delta \bf w -\bf u\delta t(\dot{\gamma}/\gamma) \\
                                   &=&  \bf u+\dot{\bf u}\delta t \\
                                   &=& \bf u (t+\delta t) 
\end{array}
$$
We have obtained thus the first interpretation.

\medskip
Notice that from (*) we also have
$$
\bf u.\partial_t(m\bf u)= m_0\dot{\gamma} 
                                   = \dot{m} 
$$
so that
$$
0= \delta t\bf u.(\bf f-\partial_tm\bf u) 
                                   = \delta E-\delta m
$$
the infinitesimal version of Einstein's law. (Recall that $c=1$).

\medskip
To obtain the second interpretation from the relativistic law of motion, we define left and right accelerations by the formulas
$$\left \{ \begin{array}{c}
\bf a_l\delta t \oplus \bf u(t)=\bf u(t+\delta t)\\ 
\bf u(t)\oplus \bf a_r\delta t=\bf u(t+\delta t)
\end{array}
\right.$$  

From these equations we obtain unique solutions
$$\left \{ \begin{array}{l}
\bf a_l=\bf a+\bf u\gamma^2( \bf a.\bf u) \\ 
\bf a_r=\gamma \bf a+\bf u(1/\bf u^2)(\gamma (\gamma -1)\;\bf a.\bf u)
\end{array}
\right.$$  
Furthermore, it is easily checked that
$$\left \{ \begin{array}{l}
\bf a_l=(1/\gamma)[\bf a_r+\bf u(1/\bf u^2)\bf a_r.\bf u(\gamma -1)] \\ 
\bf a_r=\gamma(\bf a_l+\bf u[(1/\bf u)^2\bf a_l.\bf u(\gamma -1)-\bf a_l.\bf u\gamma])
\end{array}
\right.$$  
Now, $(\delta t)_0=(1/\gamma)\delta t$ (time dilation) and thus,
$$
(1/\gamma m_0)\bf f_0=\bf a_r\;\; iff \;\;\bf f_0=\gamma \{\bf f+\bf u[(1/\bf u)^2(\bf f.\bf u (\gamma -1))-\bf f.\bf u\gamma]\}
$$
But from the law of motion $\bf f=\partial_t(m\bf u)$ (and $m=m_0\gamma$), it follows easily that $\bf f=m\bf a_l$ which in turn, implies the right hand side. But the right hand side is true, since it is precisely the Lorentz transformation of the force, as deduced from the law of motion (see the Appendix)

\medskip
To show that, conversely, the first interpretation (together with $m=m_0\gamma$) implies the relativistic law of motion, notice that from the first interpretation, $(1/m)\bf f\delta t=\bf a_l\delta t,$
which in turn implies $\bf f=m\bf a_l, $ a formula which is obviously equivalent (by taking derivatives) to  $\bf f=\partial_t(m\bf u(t),$ provided that $m=m_0\gamma.$

\medskip
Thus, we may reformulate the relativistic law of motion as
$$
\bf f=m\bf a_l
$$
where $m=m_0\gamma$ is the {\it relativistic} mass of the particle and $\bf a_l$ is {\it the relativistic} "left acceleration".

\medskip
The second interpretation, $(1/\gamma m_0)\bf f_0= a_r,$ can be written in a way that is analogous to Newton's law, namely
$$
\bf f_0=m_0\delta \bf w_0/(\delta t)_0
$$ 
(since $\delta \bf w_0=1/m_0\bf f_0(\delta t)_0$), but with the relativistic "acceleration" $\delta \bf w_0/(\delta t)_0=\gamma \bf a_r$. This expression has the following physical interpretation: assume that an observer attached to the frame $S_0$ moving with the particle let a test body `fall' freely relative to $S$. Then $-\delta \bf w_0/(\delta t)_0$ is the acceleration at the moment of the take-off, as measured by this observer in his frame (i.e. $S_0$). In classical mechanics, this acceleration is implicit in the studies of Huygens on centrifugal force ({\it De vi centrifuga}). In fact, Huygens imagines a man attached to a turning wheel and holding a thread tied to a ball of lead in his hand. The thread is suddenly cut and Huygens studies the motion of the ball at the instant when the thread is cut. This "take-off" acceleration plays an important role in some historico-critical studies such as those of the `Ecole du fil' of of  F.Reech and J.Andrade (see \cite{dug}).

\section*{Appendix:Lorentz transformations and composition of velocities}
We recall the Lorentz transformation that connects the position of an event in two inertial frames $S$ and $S'$ such that $S'$ (or rather its origin) moves with uniform velocity $\bf v$ with respect to $S.$ Indeed, if $(\bf r, t)$ describes the event in $S,$ and $(\bf r', t')$ describes the same event in $S',$ then (by mapping these quantities in an independent Euclidean 3-space):
$$ \left \{ \begin{array}{c}
\bf r'=\bf r+\bf v\gamma (v)\{(1/c^2)\gamma (v)/(1+\gamma (v)) \bf v.\bf r- t\} \\
t'=\gamma (v) (t-\bf v.\bf r/c^2)
\end{array}
\right. $$
Similarly, if a particle moves with velocity \bf u with respect to $S$ and velocity $\bf u'$ with respect to $S',$ then
$$
\bf u'= \{\bf u/\gamma (v) +\bf v [(1/c^2)\gamma (v)/ (1+\gamma (v)) \bf u.\bf v-1]\}/(1-\bf u.\bf v/c^2)
$$
Finally, if the particle is subject to a force \bf f relative to $S,$ then relative to $S'$ the force is given by
$$
\bf f'=\{\bf f/\gamma (v)+\bf v[(1/c^2)\gamma (v)/(1+\gamma (v)) \bf f.\bf v-\bf f.\bf u/c^2]\}/(1-\bf u.\bf v/c^2)
$$
These formulas may be found in \cite{rin} (pages 23, 40 and 97), although the expression $(1-1/\gamma (v))/v^2$ has been replaced by $(1/c^2)\gamma (v)/(1+\gamma (v))$ used by Ungar \cite{un}. This last formula makes sense for every $\bf v$ such that $v<c.$ 

\medskip
The Lorentz transformation of velocities may be expressed in terms of composition or addition of velocities. 

\medskip
In fact, let $V=\{\vec{v}\in {\bb{R}}^3|v^2<c^2\}.$ Define an operation on the open domain $V\times V$ of ${\bb{R}}^3\times {\bb{R}}^3$ with values in $V$ by the formula 
$$\bf v\oplus \bf u=\{\bf u/\gamma (v)+\bf v[(1/v^2)(1-1/\gamma (v))\bf u.\bf v +1]\}/(1+\bf u.\bf v/c^2)
$$
Although this operation is neither commutative nor associative, it is {\it gyrocommutative} and {\it gyroassociative}. (See \cite{un} for these notions as well as for further properties of this operation).

\medskip
In terms of this operation, we can express the Lorentz transformation of velocities of the beginning of this Appendix as
$$
\bf u'=(-\bf v)\oplus \bf u
$$
or, equivalently, as
$$
\bf u=\bf v\oplus \bf u'
$$
Notice that even if $\bf u$ and $\bf v$ are velocities, $\bf v\oplus \bf u$ is not interpretable as a velocity. To make contact with actual composition of velocities we need to devise physical set-ups to realize each vector as a velocity (relative to a suitable frame) and the operation as actual composition of velocities according to our diagram
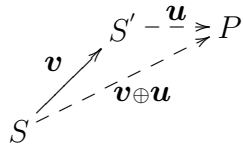
\begin{center}
\leavevmode
\xy\xymatrix{
& S^{'} \arrow @{-->} [r]^{\bf u}  &P \\
S \arrow @{->} [ru]^{\bf v} \arrow @{-->} [rru]_{\bf v \oplus \bf u} &\\
}
\endxy
\end{center}

\section*{Acknowledgments}
The first author owes a great debt of gratitude to Nicanor Parra. He formulated clearly the problem of the status of the 3-force and the law of motion in the Theory of Special Relativity and told him about the take-off acceleration in classical mechanics, suggesting that it could be used in Special Relativity. He had countless conversations on and off on  `natural and supernatural' mechanics, since he followed Parra's course in 1957 at the Instituto Pedag\'{o}gico of the Universidad de Chile in Santiago. His encouragement is greatly appreciated. He is  also in debt to Jorge Krausse. Besides discussions of a general nature, he helped him to find his way in the literature on Relativity Theory.

\end{document}